PeBLes: Prediction of B-cell epitope using molecular layers

Naga Bhushana Rao .K[1], Ranjit Prasad Bahadur[2*]

1) Advanced Technology Development Center, Indian Institute of Technology Kharagpur, Kharagpur-721302, West Bengal, India.

2) Computational Structural Biology Lab, Department of Biotechnology, Indian Institute of Technology Kharagpur, Kharagpur-721302, West Bengal, India.

*To whom correspondence may be addressed
E-mail: r.bahadur@hijli.iitkgp.ernet.in
or
ranjitp_bahadur@yahoo.com

Phone: +91-3222-283790
Fax: +91-3222-27870

**Abstract**

Characterization of B-cell protein epitope and developing critical parameters for its identification is one of the long standing interests. Using Layers algorithm, we introduced the concept of anchor residues to identify epitope. We have shown that majority of the epitope is composed of anchor residues and have significant bias in epitope for these residues. We optimized the search space reduction for epitope identification. We used Layers to non-randomly sample the antigen surface reducing the molecular surface to an average of 75 residues while preserving 50% of the epitope in the sample surface. To facilitate the comparison of favorite methods of researchers we compared the popular techniques used to identify epitope with their sampling performance and evaluation. We proposed an optimum $Sr$ of 16 Å to sample the antigen molecules to reduce the search space, in which epitope is identified using buried surface area method. We used the combinations of molecular surface sampling, anchor residue intensity in surface, secondary structure and sequence information to predict epitope at an accuracy of 89%. A web application is made available at http://www.csb.iitkgp.ernet.in/applications/b_cell_epitope_pred/main.

**Introduction**

Antigen-antibody interactions play potential roles in defending human from various pathogens and threats (2, 3). Antibodies are protein molecules secreted by B lymphocytes to neutralize the antigen molecules. Typical antigen-antibody molecules shows very low $k_d$ that can be attributed to the high affinity and avidity of antibody molecules (5). Moreover, the affinity of antibodies to specific antigens can be modulated through their engineering (4). Antibodies and most protein antigens follow the norms of protein folding that are generally obeyed by structured proteins, while some antigens are disordered (6). Epitope is the surface of antigen that binds antibody, whereas, paratope is the interacting surface of antibody (8). Epitope can be classified into conformational and linear epitopes based on their continuity in the peptide sequence. Conformational epitopes are formed from the residues that have spatial proximity but are distant from each other in primary sequence. Whereas linear epitopes refer to the continuous stretch of residues in the polypeptide (9, 10). Epitopes are also classified into structural and functional based on their role in binding with the antibodies (11).

B cell epitopes have specific characteristics, which can be used to discriminate them from the non-interacting surface of proteins (12). Based on the characteristic features of epitope, different prediction models were developed to predict them from the linear sequence as well as from the tertiary structures (REF). Performance of the prediction models have wide variations in their accuracies (13–15). A complete understanding of epitope determining features is still limited (16), which further resulted in  poor performance of the prediction models (15). The problem lies in the understanding of the antigenicity of a protein (19, 20). In biological systems any protein could be an antigen, provided the host organism is different or sometimes even within the same species. A single protein molecule may show multiple epitopes with respect to the host  organism (21). Molecular complexity and screening mechanism of host organisms result in variations in epitope selection (22, 23). One such major reason is positive and negative screening of antibodies (19).

Availability of structural data of antigen antibody complexes facilitated the precise identification of epitope, which further helped their characterization (24, 25). Epitope identification with structural complexes used two major types of methods: one is based on distance between the interacting atom pairs and the other based on the amount of solvent accessible surface area (SASA) buried upon interaction. Being at the molecular surface, search space to identify the epitope can be reduced by working with the surface instead of the entire molecule.

We recently develop Layers, which extracts the molecular surface besides non-random sampling of the surface (28). It has long been established that epitope is part of protruded areas (29). Non-random sampling using Layers is shown to preserve protrusions at the molecular surface. We reduce the search space for epitope prediction by using only sampled surface layers extracted using Layers (30). We proposed a optimu Sr for surface sampling for epitope identification. Besides, we also introduced the concept of anchored residues, which spans from the surfaced layers to its underneath layers. We find these anchor residues play significant role at the epitope, and can be used efficiently to identify the B-cell epitope. We have used deep learning method to predict the epitope from a given tertiary structure with an accuracy of 89%.

**Results**

**Analysis of Epitope mapping and statistics on antigen-antibody interfaces**

Interface of an antigen-antibody complex is identified using two different methods: one is based on Euclidian distance with two different cutoffs (4.0 Å and 4.5 Å) and another is based on Solvent Accessible Surface Area (SASA). Typically, molecular geometry of a protein antigen can be classified into three areas: interacting surface, non-interacting surface and the interior that is not directly exposed to the environment. Molecular surface of the antigens is extracted using the Layers algorithm (28) and the non-interacting surface of antigen is identified by removing the atoms of epitope from the surface layer.

Table1 shows that the antigens are generally smaller in size compared to antibodies. This is also reflected at their interfaces, where antibody contributes more to the BSA than antigen.

. Antibody is consistently and significantly bigger than antigen in terms of atoms and residues. It is interesting to see that this trend is not observed in epitope and paratope sizes. In all three methods the paratope is at least 9% larger than the epitope in terms of residues or atoms, implying that paratope wraps the epitope completely (Table 1).

Epitopes identified using SASA typically are larger than the epitopes identified using distance based methods.

**Propensity of amino acid residues in epitope**

The propensity of amino acid residues at the epitope and at the protein surface is shown in Figure 1. Residues that are significantly preferred in epitope are Tyr, Pro, His, Asn, Gln, Lys, Arg, Asp which have the propensity more than 0.2 in at least one of the methods. Residues, Gly, Thr, Ser, Trp, Glu also prefer to present at the epitope, however their propensity is very low (less than 0.2). Six residues, Ile, Val, Leu, Phe, Cys, Ala show strong negative propensity signifying that they are not preferred in epitope. This high intensity of negative propensities of these residues may help to discard the surface patches of a protein reducing the search space for epitope identification.

Residues Trp and Met shows some interesting propensities. In general, Trp is preferred at the protein surface and also in the epitope identified by D2 and BSA methods. However, it shows negative propensity at epitope by D1 method. In contrary, Met is not preferred at protein surface, which is also reflected in epitope identified by D2 and by BSA methods. However, it shows very low positive propensity in epitope using D1 method. The propensities obtained for epitope calculated with BSA correlates better with D2 than with D1. Residues Ser, Gly show interesting propensities: preferred at the epitope, however, depleted at the protein surface. The other three residues, Thr, Phe, Leu, have zero propensities at the protein surface, however, Thr is preferred at the epitope and Phe and Leu are strongly depleted from the epitope.

**Layers sampling and epitope retention**

Epitope is a part of molecular surface, and often belongs to the protruding regions of the structure (29). Hence, it is implied that search space to identify epitope can be reduced by working with molecular surface instead of the entire molecule. Further reduction in search space can be achieved by using a method that can retain global protrusions while losing the local ones (28). However, surface sampling to generate coarse representation of a molecule should be optimized such that the crucial information or rather epitope is preserved besides reducing the molecular surface. We lack firsthand information of atoms or residues that are crucial and should be preserved while sampling. Hence, non-random sampling of surface is preferred to random sampling because of its reproducibility and uniformity across the molecules. We used Layers (28) for non-random sampling of molecular surface, which also has the ability to fine-tune the coarse representation. Our aim is to obtain an optimum $Sr$ value for sampling the molecular surface which gives optimum surface reduction and epitope retention.

The atoms or residues retained in the sampled surface belong to either epitope or to the non-interacting surface of the antigen. The epitope identified with three different methods varies in their size, shape and composition. Hence, we compared the ability of sampling to retain epitope in their respective methods. Figure 2 shows the retention of epitopes with non-random sampling in three different methods (D1, D2 and BSA). We find that both molecular surface and the epitope retention decreases with increased $Sr$ (Fig 2, 3). A minor variation in sampling is observed between the structures as well as between the methods used to identify epitopes. Distance based methods show an overlap in the higher retention and lower retention zones, making it difficult to pick an optimum $Sr$ for sampling. Moreover, no significant variation is observed between D1 and D2. (Fig. 2a, b). However, the SASA method generates a clear distinction between the higher and the lower retention zones of epitope (Fig. 2c). The progressive transition of epitope retention in BSA method facilitates to select the optimum $Sr$ for non-random sampling. Figure 3 shows an exponential reduction in molecular surface up to $Sr = 16$ Å using non-random sampling. Accordingly, we have used $Sr = 16$ Å, which retains 50% of the epitope and only one third of the entire molecule.

We should ensure that non-random sampling is free from any bias in removing the surface atoms. The result could be an affirmation that the non-random sampling is uniformly removing atoms from the surface. Figure 4 shows the percentage of residues belongs to epitope in coarse structures. Non-random sampling with Layers (28) should preserve the occupancy of epitope in sampled structure and should be close to constant among all the coarse structures irrespective of $Sr$. This means that the atoms from the surface are removed evenly from both non-interacting surface and from epitope without any bias. Figure 4 shows that the epitope constitutes only 10% of the molecule. Compared to native structures epitope occupancy in sampled structures increases by 2 to 8%, which is expected as sampling loses the protein interior. It is interesting to notice that the occupancy of epitope in the coarse structures remain constant even with increased $Sr$. This constant occupancy of epitope in coarse structures can be attributed to the non-random sampling. Sampling with $Sr$ beyond 16 Å results in relatively lower reduction in molecular surface besides loosing higher percentage of epitope. However, epitope is always retained in any coarse structure obtained with $Sr$ upto 30 Å (Fig. 4).

**Anchor residues in epitope**

We introduced Residue Transition Pattern (RTP) in Layers (28), which is a theoretical model to represent the residue position in a folded protein structure. According to Layers, residues are assigned to a particular layer from inside to outside of a molecule. We find many residues extend across multiple layers in a molecule. If any atom of such multi-layer residues is at the molecular surface, we assign them as anchor residues. Anchor residues experience more conformational constraints compared to other residues as they are buried in multiple layers. This makes them almost rigidly placed on protein surface and their spatial arrangement may create a unique surface patch for molecular recognition.

We explored the presence of anchor residues at the molecular surface. Using BSA, we identified 2193 epitope residues of which a significant number (93 %) are anchor residues (Table 2). We find significantly high (~0.5) positive propensity of anchor residues at the epitope identified by all the three methods. This suggests that the anchor residues are strongly preferred at the epitope compared to non-interacting surface. Continuous stretch of residues with three or more in epitope was checked for their anchored nature. We find 272 such continuous stretches in the dataset. Of these stretches, we identified 95 are anchored on both N and C terminals, 223 are anchored only on N terminal and 210 are anchored only on C terminal (Table 2). Besides residue stretches, we have also look for singlet anchor residues. Out of 2046 anchored residues, we identified 439 singlet anchored residues. Similar results were obtained for D1 and D2 methods also (Table 2).

**Epitope prediction**

Deep learning (DL) and deep belief networks mimic the artificial neural networks inspired from the biological neural networks and are one of the state of the art representation learning models that are excelling in image processing and machine learning (31, 32). Various other machine learning models were used till date to predict the B cell epitopes (13, 33, 34). We opted to use DL because of its self-learning abilities bundled with extremely fast and massive scalable frameworks. In order to feed the epitope data to the DL frameworks we have to carry out some transformation operations on epitope data to make it compatible with the framework. Besides transforming the structural data of epitope, we should also ensure the incorporation of significant features like anchored residues, secondary structure composition, sequence information…, into the framework.

First, we transformed the epitopes along their principal axis, and aligned all the transformed epitopes to the *yz* plane using VMD (Fig. 5) (35). A two-dimensional grid is taken as a reference, which is also aligned to *yz* plane. Each grid point is placed 1 Å apart from the neighboring grid point (Fig. 5b). Epitopes are mapped to this reference grid and the absolute value of *x* coordinate of the atoms is assigned to a grid cell (Fig. 5b). Four grid points will make up a cell to which the information is fed. The *y* and *z* coordinates of the atoms in epitope are used to map the atoms into their suitable cell in the grid. If multiple atoms are mapped to a particular grid cell, then the values of *x* coordinates are summed with existing values and will be reassigned to the cell. Mapping of epitope to the grid is followed by the normalization of the values assigned to each cell. Head block (group of grid cells) of the grid is filled with the parameters including features based on anchor residues, secondary structures, structure and size of patch, number of residues and amino acids composition in patch.

Grids mapped with epitope represent the positive instances for training. For negative instances, we generated non-overlapping surface patches from non-interacting surfaces of antigens. These raw patches were checked for their compliance with the epitope radius. A raw patch is selected for training only if its radius is within the average epitope radius ± the standard deviation. Table 3 shows the number of raw and selected training patches. These selected patches were also transformed and aligned to the origin and *yz* plane as is done with epitopes. These transformed and aligned patches constitute our negative dataset.

Training with DL is carried out for 10 iterations resulting in prediction accuracies as shown in Table 3. We stopped training after 10 iterations as the error rate of learning did not diminish further. We have obtained high prediction accuracy (89 %) using the BSA method (Table 3). D2 method also give significant prediction accuracy (80 %), however, D1 gives mediocre accuracy (68%). We have developed a web application for the B-cell epitope prediction called PeBLes, which is available free at http://www.csb.iitkgp.ernet.in/applications/b_cell_epitope_pred/main.

**Discussion**

Parasites coevolve with their hosts resulting in resistant strains. This challenges the existing techniques to fight against them. This demands new strategies to combat the evolution of parasites (36, 37) (38). Genome information of parasites can directly help to predict the epitope, potentially bypassing the structural dependency for epitope prediction (39). Moreover, some antigenic peptides does not have a defined three dimensional structures (6). In order to develop genome-based prediction model, we need to have more reliable abstractions to identify epitope. Computational tools and models for epitope prediction, once standardized, can be more promising and safer for genome-wide epitope identification (40).

The functional diversity of proteins is encrypted in their amino acid composition and this applies to the epitopes also. Propensity of amino acid residues in epitopes is an important metric to deduce an unbiased preference. Residue propensity in epitope shows significant implications, which correlates with the specific role of residues in protein structures. Jens et al., (41) found positive propensity for Trp in epitope. Our findings are in accordance with them except for the D1 method, which shows very low negative propensity for Trp (Fig. 1). The minor variation may be due to the removal of six structures from the original dataset besides change in definition of antigen surface using Layers. Besides Trp, propensity of Met derived using D1 also shows the trend that is against the trend observed in D2 and BSA methods. This may be due to the fact that the D1 methods narrowed down the surface patch of epitopes, resulting in change in residue composition. On the other hand, epitope determined using D2 broaden the surface patch, which correlates well with that obtained through BSA methods.

The set of residues with negative propensities (Fig. 1) are known to have distinct functions in protein structures. One of them is Cys, which has the able to form disulphide bridges (42), most often occurring at the protein interior or across the two polypeptides that eventually gets buried leaving little scope to be accessible at the molecular surface. The other residues, which are hydrophobic in nature, are often buried in protein structures and occasionally found on the surface (43). Hence, their propensity at epitope is also very less.

One of the striking characteristics of epitope is their occurrence at the protruding part of a molecule (29). However, defining protrusion in an abstract context, which is often used in some algorithms, has its own limitations (44). Proteins are arbitrarily shaped molecules, and protrusions cannot be generalized with a predefined geometric shape (44). Layers, on the other hand, do not generalize the shape of the molecule. Layers perform non-random sampling of molecular surface with respect to its shape preserving the global protrusions (28). It is evident from this study that the epitope, be it linear or conformational, belongs to the protruding regions of the protein molecule (Fig. 2, 4). This feature of epitope gives us a leverage to reduce the search space and accurately predict them. Figure 4 clearly shows that the non-random sampling maintained a constant occupancy of epitope in coarse structures; however, it significantly reduces the molecular surface. Thus global protrusions and non-random sampling of molecular surfaces can be efficiently used for epitope identification while reducing the search space.

The default value of $Sr$ used in Layers is 1.52 Å that extracts the entire surface of the molecule, which is expected to retain the entire epitope. However, in some instances, it may happen that the underneath layer of the surface layer may also be tagged as epitope because of its spatial proximity with the antibody. This second layer of atoms, which are labeled as epitope, may not be extracted as surface layer (Fig. 2). However, this loss is very negligible and can be ignored. The average number of atoms and residues in epitope and paratope suggests that the latter is at least 9% larger than the former, suggesting that antibody can efficiently capture the antigen (Table 1). The excess of residues or atoms seen in antibody structures may serve different purpose; binding the cells that clears the antigen-antibody complexes from host system (45). Sampling with Layers results in significant reduction for large proteins and negligible reduction for small proteins although keeping the overall molecular shape in both cases (28). This observation is found in sampling with antigens (Fig. 3). For the large antigens, the number of residues retained in surface sampling shows exponential fall; however, for small antigens, the number of residues retained remains almost constant with increased $Sr$ (Fig. 3). Moreover, the non-random sampling shows that the epitope and non-interacting surfaces were sampled without any bias (Fig. 4).

F1 score is a performance measure calculated using precision and the recall abilities of a prediction model over the test cases. For all three methods we obtained almost same F1 score their respective prediction models. This F1 score represents the trade-off between precision and recall. Since F1 score is high it represents that both precision and recall are also high.

Unlike the non-overlapping patches used in training, we use overlapping patches of molecular surface for the prediction of B-cell epitopes. This is justified by the fact that the overlapping patches results in large number of negative instances compared to positive instances. This may introduce bias in learning, which may further results in poor prediction model. For a query structure submitted to the webserver, it will use overlapping patches for prediction. Prediction results on the overlapping patches will be checked for the consensus and an epitope will be labeled and given back as prediction result.

**Conclusion**

B-cell epitope prediction with reasonable accuracy is one of the long awaited computational tool. Owing to the fact that epitope belongs to protruding regions of proteins we used Layers algorithm to sample the antigen molecules. We have proposed 16 Å as an optimum $Sr$ for the antigen-antibody dataset where molecular surface reduction and epitope retention are higher. We introduced the concept of anchor residues which have significant occupancy in the epitopes. A positive propensity of 0.44 is obtained for the propensity of anchor residues in epitope. Combination of molecular surface sampling and anchor residue intensity in surface can help determining the location of epitope more reliably. Usage of such significant factors resulted in a high accuracy prediction model with 89% prediction accuracy.

**Materials and Methods**

**Dataset of antigen-antibody complexes**

The dataset of antigen-antibody complexes for the B-cell epitope is taken from Jens et al, (41) which reports 107 structures. We have discarded six complexes (PDB id: 1N0X, 2OSL, 1QGC, 2HFG, 1NL0, 2J4W) from the original dataset that have less than 40 residues. Secondary structure assignments were performed using the program DSSP (46, 47).

**Identification of epitope**

Epitope is identified using two different methods: one using Solvent Accessible Surface Area (SASA) and the other using Eucledian distance. The buried surface area (BSA) (30).between antigen and antibody is defined as the sum of the SASAs of the individual subunits less that of the complex. All the residues and their corresponding atoms that lose SASA upon complexation are considered as interface. SASA values are calculated using the program NACCESS (26), which implements the Lee and Richards algorithm (27). In distance based method, interface atoms are identified using two different cut-off distances of 4.0 Å named as D1 and 4.5 Å named as D2 methods.

**Propensities of residues**

Propensities of amino acids residues in epitope are calculated using the equation 1.

Propensity of residue type **X** at epitope

$$= ln \left[ \frac{\frac{Number\ of\ residues\ of\ type\ \textbf{X}\ at\ epitope}{Total\ residues\ at\ epitope}}{\frac{Number\ of\ residues\ of\ type\ \textbf{X}\ at\ surface}{Total\ residues\ at\ surface}} \right] \quad ...\ eq.\,1$$

The propensity of amino acid residues in protein structures were calculated using 14,221 non-redundant structures taken from Pisces (48) using equation 2. This dataset consists of structures having less than 35 % sequence identity between any two polypeptide chains, with resolution better than 3.5 Å and have sequence length ranges between 40 and 10000 residues.

Propensity of anchored residues at epitope

$$= ln \left[ \frac{\frac{Number\ of\ anchored\ residues\ at\ epitope}{Total\ residues\ at\ epitope}}{\frac{Number\ of\ anchored\ residues\ at\ surface}{Total\ residues\ at\ surface}} \right] \quad ...\ eq.\,2$$

We define Anchor residues as those at the surface layers yet spans through at least one inner layer below the surface. The propensities of Anchor residues in epitope are calculated using the equation 3.

Propensity of anchored residues at surface

$$= ln \left[ \frac{\frac{Number\ of\ anchored\ residues\ at\ surface}{Total\ residues\ at\ surface}}{\frac{Number\ of\ residues\ at\ surface}{Total\ residues\ in\ protein}} \right] \qquad ...\ \boldsymbol{eq.3}$$

**Surface patches**

Surface patches were defined using two different approaches. In one, we used the atoms at surface layers to define the surface patch, in another we used center of geometry (CG) of residues at non-interacting surface layers. CG of a residue is calculated using only those atoms of that residue that are present in the surface layer. For a given residue at the surface layer, we define a surface patch with all its neighbors whose CG are within the average radius of the epitope (Table 3). We have used three different cutoff values, two based on distance and one based on SASA, for defining the surface patches.

**Surface patches used for training and testing**

We have used epitopes as positive dataset for training and testing. For negative dataset we used the non-interacting surface patches of antigens, which are non-overlapping. A surface patch is considered as valid patch if its radius is within the average epitope radius including the standard deviation. These valid patches were used for training and testing the model. The nolearn package of python is used to train and predict the eptiopes (49).

TP True positive

TN True negative

FP False positive

FN False negative

$$precision = \frac{TP}{TP + FP}$$

$$recall = \frac{TP}{TP + FN}$$

$$F1 = 2 \cdot \frac{precision \cdot recall}{precision + recall}$$

**Conflict of Interest**

The author dclares no conflict of interest.

**Authors' Contribution**

NBRK and RPB conceived the idea, worked on the problem and wrote the manuscript together.

**Figure Legends**

**Figure 1:** Propensity of amino acids in epitope identified using different methods compared with the general trend of amino acids propensities in proteins in general.

**Figure 2:** Retention of percentage of epitope residues by sampling the antigen surface with Layers with the following methods. (a) D1. (b) D2. (c) BSA.

**Figure 3:** Average, minimum and maximum number of residues retained in antigen molecules by surface extraction and sampling using Layers with $Sr$ of Layers varied from default 1.52 Å to 30 Å.

**Figure 4:** Average, minimum and maximum percentage of epitope retained in antigen with Layers extraction and sampling by varying $Sr$ of Layers from default 1.52 Å to 30 Å. Three different methods used to identify epitope are compared.

**Figure 5:** Transformation and alignment of epitopes to the origin in three-dimensional space and mapping the epitope data to a two-dimensional grid. (a) Epitopes as is in the three-dimensional space before transformation and alignment. (b) Side view. (c) Top view.

**Figure 1**

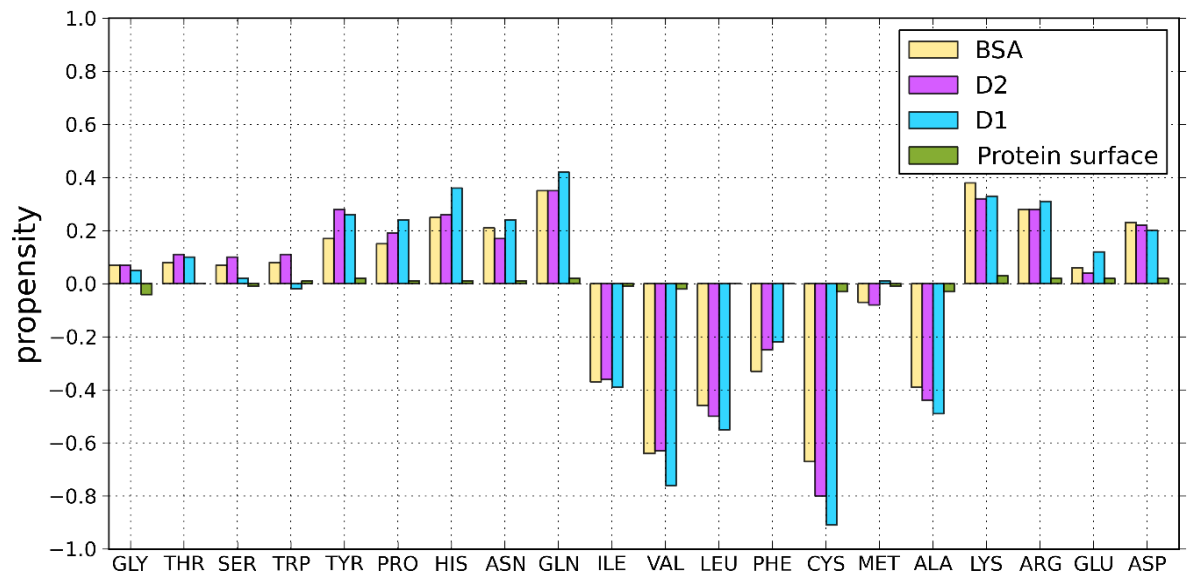

**Figure 2**

**a**

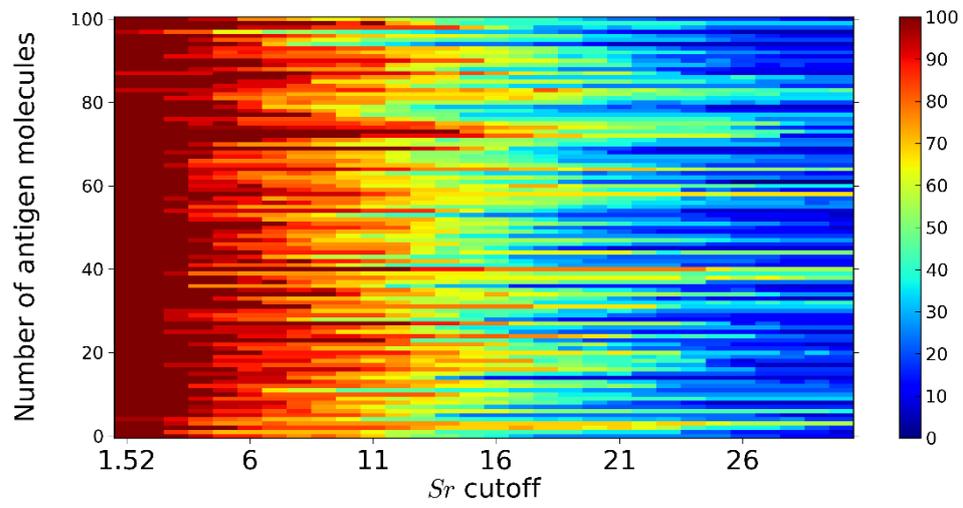

**b**

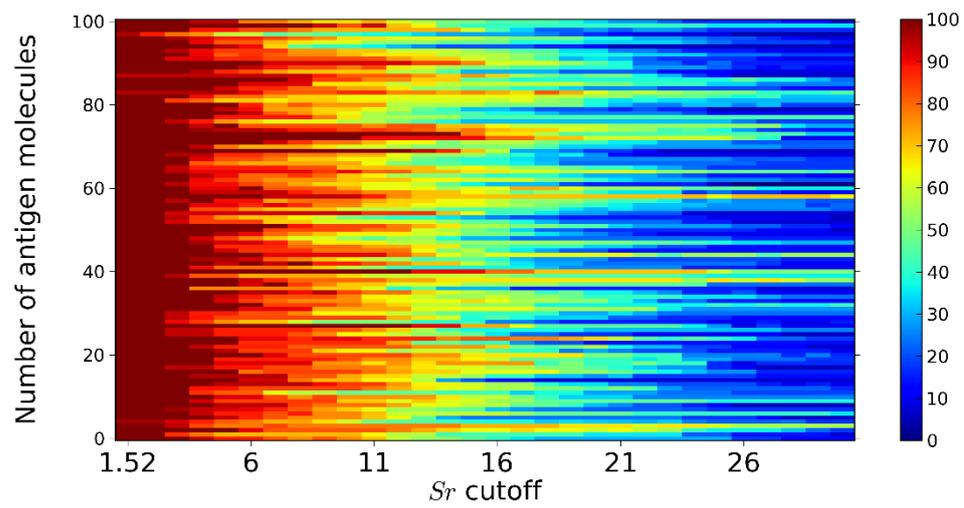

**c**

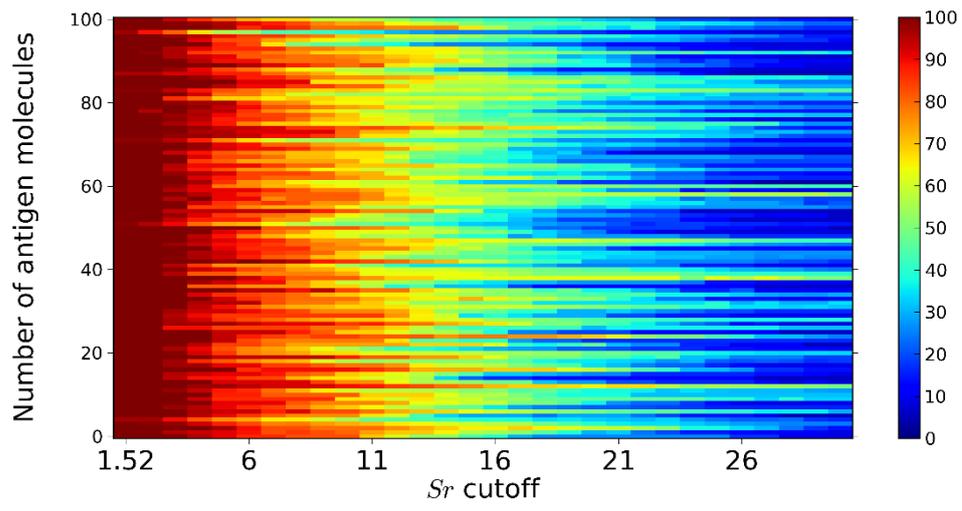

**Figure 3**

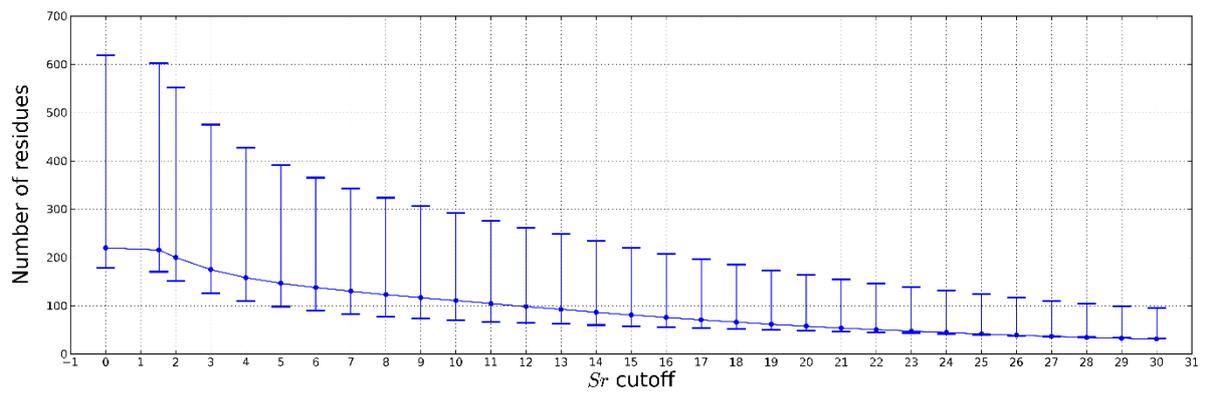

**Figure 4**

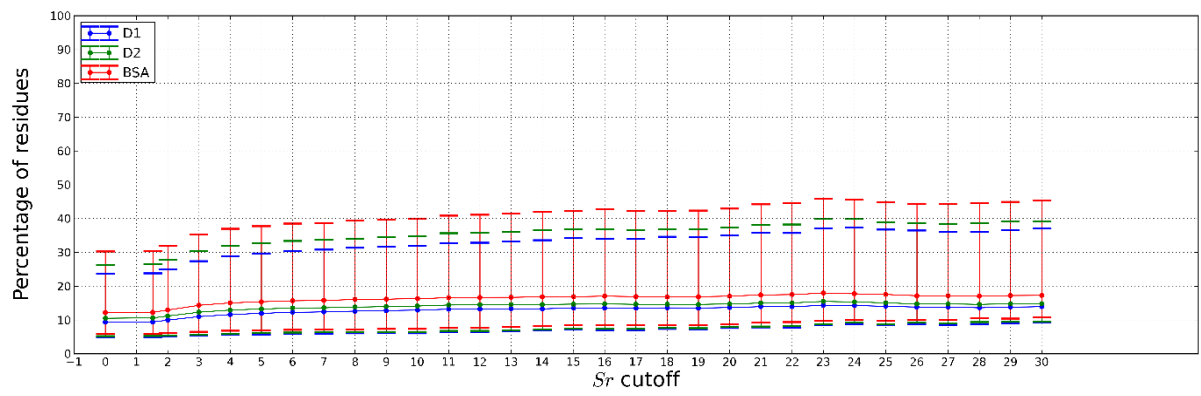

**Figure 5**

a

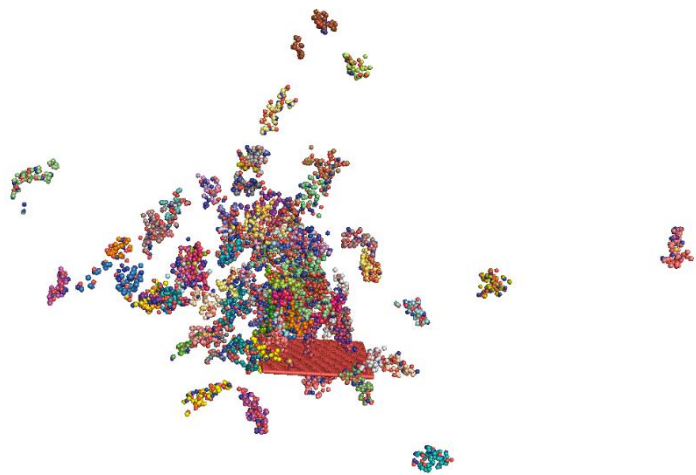

b

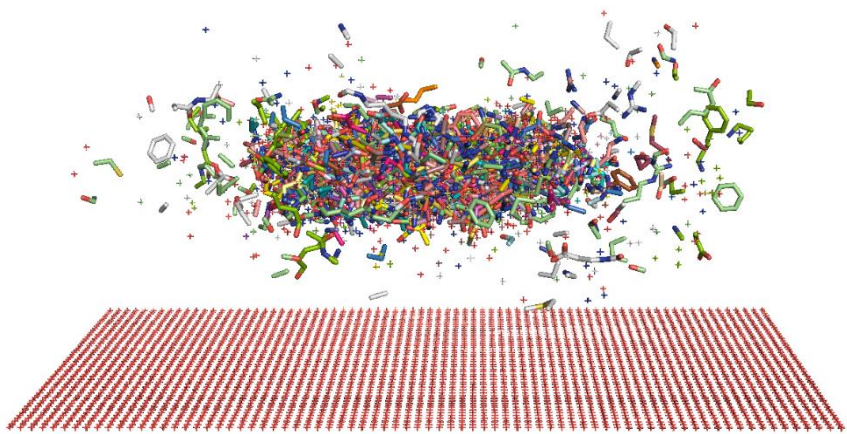

c

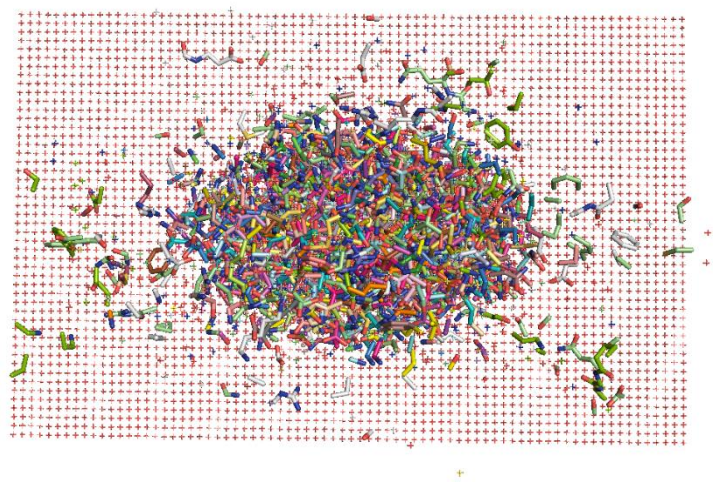

**Table 1: Average (standard deviation) of antigen antibody interface:**

| | D1 | D2 | BSA |
|---|---|---|---|
| Number of atoms | | | |
| Antigen | 1820.90 (±1407.38) | | |
| Antibody | 3081.04 (±530.77) | | |
| Number of residues | | | |
| Antigen | 231.43 (±176.61) | | |
| Antibody | 399.76 (±68.81) | | |
| Number of atoms in interface | | | |
| Antigen | 54.80 (±19.58) | 78.58 (±30.04) | 71.24 (±28.75) |
| Antibody | 59.14 (±19.28 | 87.42 (±29.50) | 77.34 (±29.92) |
| Number of residues in interface | | | |
| Antigen | 16.45 (±6.78) | 18.70 (±8.02) | 21.74 (±9.32) |
| Antibody | 17.93 (±6.20) | 20.88 (±7.63) | 23.86 (±9.35) |
| SASA | | | |
| Antigen | — NA— | — NA— | 413.04 (±165.38) |
| Antibody | — NA— | — NA— | 408.02 (±186.12) |

**Table 2: Anchor residues in epitope**

|  | D1 | D2 | BSA |
|---|---|---|---|
| Total residues | 1658 | 1886 | 2193 |
| Total Anchor residues | 1480 | 1748 | 2046 |
| Number of continuous stretches | 207 | 245 | 272 |
| Both N, C terminals anchored | 69 | 74 | 95 |
| N-terminal anchors | 155 | 181 | 223 |
| C terminal anchors | 154 | 171 | 210 |
| Singlet anchors | 397 | 384 | 439 |
| Anchor residues in antigen surface | 12581 | 12448 | 12297 |
| Total surface residues in antigen | 22921 | | |
| Total residues in antigen | 23374 | | |
| Propensity of anchor residues | 0.44 | 0.48 | 0.49 |

N-anchor: Continuous stretch of epitope residues starting with an anchor residue on N-terminal side of that stretch. Ignored if the stretch is only one residue long.

C-anchor: Continuous stretch of epitope residues ending with an anchor residue on C-terminal side of that stretch

**Table 3: Mean and standard deviation of radius of epitope**

|  | **D1** | **D2** | **BSA** |
|---|---|---|---|
| Mean ($\pm$ Standard deviation) of epitope radius | 8.47 ($\pm$ 2.30) | 8.68 ($\pm$ 2.31) | 9.32 ($\pm$ 2.34) |
| Number of raw patches | 4307 | 4136 | 3642 |
| Number of selected patches | 740 | 549 | 220 |
| Prediction accuracy | 68 | 80 | 89 |
| F1-Score | 95 | 96 | 96 |

Non interacting surface raw patches with residue pairs considered only once and valid patched identified by radius of patch within the epitope radius (radius-std $\leq$ X $\leq$ radius+std)